\documentclass[fleqn]{annalen}

\usepackage{epsfig}

\begin{document}


\newcommand{\volume}{8}              
\newcommand{\xyear}{1999}            
\newcommand{\issue}{Spec. Issue}               
\newcommand{\recdate}{29 July 1999}  
\newcommand{\revdate}{9 August 1999}    
\newcommand{\revnum}{1}              
\newcommand{\accdate}{20 August 1999}    
\newcommand{\coeditor}{M. Schreiber}           
\newcommand{\firstpage}{237}         
\newcommand{\lastpage}{240}          
\setcounter{page}{\firstpage}        



\newcommand{\keywords}{disordered solids, metal-insulator transitions} 


\newcommand{\PACS}{61.43.--j, 71.30.+h, 73.40.Hm}


\newcommand{\shorttitle}{V. Senz et al., Metal-insulator transition in 
system with easy spin axis} 


\title{Metal-insulator transition in a 2-dimensional system with 
an easy spin axis}

\author{V. Senz,$^{1}$ U. D\"{o}tsch,$^{2}$ U. Gennser,$^{2}$ 
T. Ihn,$^{1}$ T. Heinzel,$^{1}$ K. Ensslin,$^{1}$\\ R. 
Hartmann,$^{2}$ and
D. Gr\"{u}tzmacher.$^{2}$}

\newcommand{\address}
  {$^{1}$ Solid State Physics Lab., ETH Z\"{u}rich, CH-8093 
Z\"urich, Switzerland\\
$^{2}$ Paul Scherrer Institute,  CH-5232 Villigen PSI, Switzerland }

\newcommand{\email}{\tt ihn@solid.phys.ethz.ch} 
\maketitle

\begin{abstract}
The low-temperature resistivity of a SiGe 2-dimensional hole gas has 
been
studied using the gate controlled carrier density as a parameter.  A
metal-insulator transition is seen in both the temperature and the 
electric
field behaviour. Values of 1 for the
dynamical exponent and 2.85 for the correlation length exponent
are obtained from scaling plots. No quenching of the metallic phase 
in a parallel
magnetic field is observed. Since in our system there is an easy axis 
for
magnetization, this result supports the hypothesis that the 
transition is
related to spin interactions.
\end{abstract}


The recent discovery of a metallic phase at zero magnetic field in
two-dimensional systems \cite{Krav1} challenges the theoretical 
considerations
predicting a localized phase for these systems.  There have been 
reports on MIT in
Si-MOSFETs \cite{Krav1}, p-SiGe \cite{Cole}, p-Ga[Al]As \cite{Simm}, 
high
density n-Ga[Al]As with self-assembled InAs quantum dots \cite{Ribe}, 
and
recently also indications of a transition in n-Ga[Al]As with 
extremely low
carrier concentration \cite{Hane}.    Here, we present evidence for 
temperature scaling
and electric field scaling at densities around the MIT in a SiGe 
2-dimensional
hole gas (2 DHG), and obtain values for the two scaling coefficients.

The samples employed in this study were grown by 
molecular beam epitaxy, and
consist of a 200{\AA} $\rm{Si_{0.85}Ge_{0.15}}$ layer surrounded by 
undoped Si
layers, a 150{\AA} B-doped Si layer with a setback of 180{\AA} from 
the
well, and a 200{\AA} undoped Si cap.  The SiGe layer forms a 
triangular
potential well for the 2 DHG (Fig. 1a).   Due to the lattice mismatch 
between
the Si and the SiGe and due to size quantization, the heavy hole 
($m_J$ = $\pm$3/2) potential is split from
the light hole ($m_J$ = $\pm$1/2) potential, and ensures that the 
lowest bound
state has heavy hole character. The transport effective mass of
this state is $m^* \approx 0.25 m_0$, as extracted from the 
temperature
dependence of Shubnikov -- de Haas oscillations.  Conventional Hall 
bar
structures were fabricated with the typical length between voltage 
probes of the
order of 10 $\mu$m, and the length between current contacts and width 
of the
Hall bars 60 -- 140 $\mu$m and 10 -- 30 $\mu$m, respectively. The 
carrier
density $n_s$ could be tuned, $1.8\cdot 10^{11}\; \mbox{cm}^{-2} \le  
n_s  \le
2.6\cdot 10^{11}\; \mbox{cm}^{-2}$, using a Ti/Al Schottky gate. The 
mobility in these
structures was found to increase strongly with carrier concentration, 
from 1000
$\mbox{cm}^2/Vs$ (for $n_s  = 1.8\cdot 10^{11}\; \mbox{cm}^{-2}$) to 
7000 $\mbox{cm}^2/Vs$ ($n_s 
= 2.6\cdot 10^{11}\; \mbox{cm}^{-2}$). 
\begin{figure}[t] 

\epsfig{file=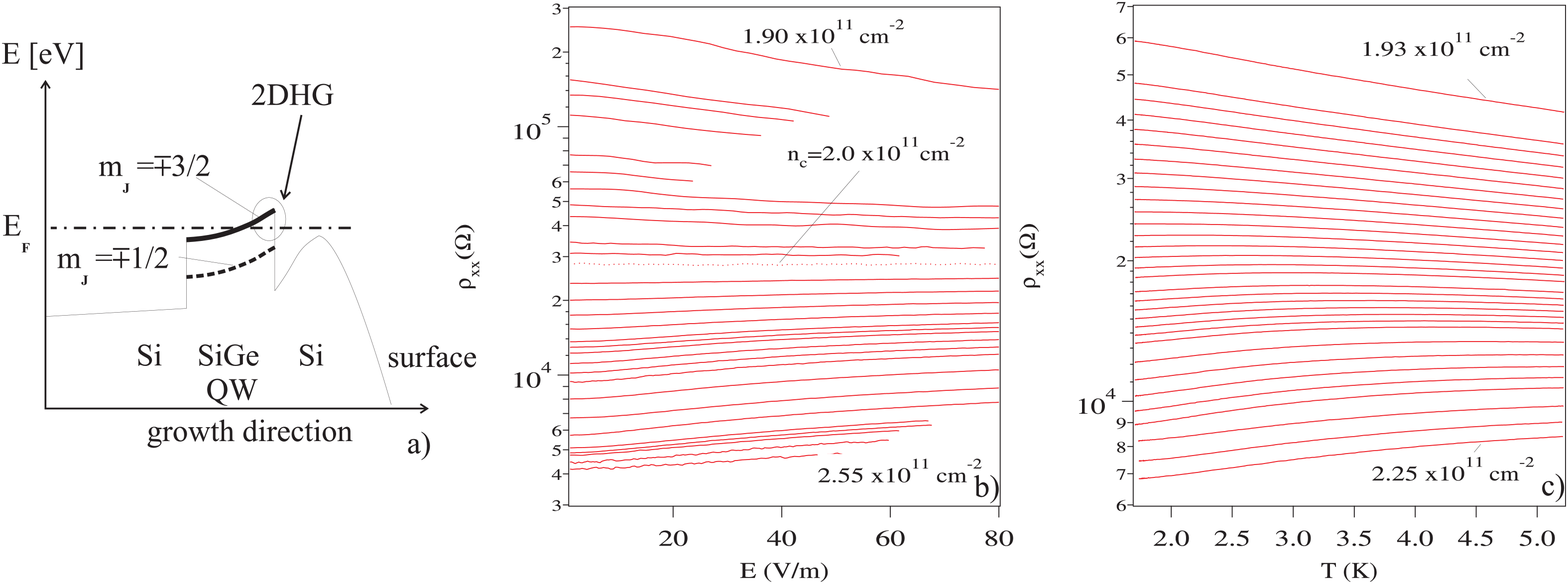,width=\linewidth}
\caption{a):Schematic diagram of the valence band structure of the 
sample.
The heavy hole potential ($m_J = \pm 3/2$; full line) and light hole 
potential
($m_J = \pm 1/2$, dashed line) are separated in the SiGe well due to 
the built-in
strain. The $m_J = 3/2$ lies partly above the Fermi energy 
(dashed-dotted line), and
the 2 DHG is formed at the interface between the SiGe QW and the top 
Si layer.
b):  Resistivity as a function of electric field E for carrier 
densities in the
range $(1.90 - 2.55)\cdot 10^{11}\; \mbox{cm}^{-2}$.  The critical 
resistivity is marked with a dotted line.
c): Resistivity as a function of temperature T
for carrier densities in the range $(1.93 - 2.25)\cdot 10^{11}\; 
\mbox{cm}^{-2}$.}
\end{figure}

Current-voltage characteristics were measured with a DC-technique at $30\; mK$.
In Fig. 1b the differential resistivity $\rho_{xx}$ is plotted as a function of 
electric field for
different carrier concentrations. Plotting the differential resistivity 
makes undesired DC-offsets caused by our amplifier disappear and 
allows us to compare directly the different curves.  At high carrier densities (low 
$\rho_{xx}$), the
behaviour is that of a metal: for large electric fields the 
resistivity increases.  For
the lowest carrier concentrations, the behaviour is instead 
insulating, with a
lower resistance at higher E-fields.  At a critical density $n_c = 
2.0\cdot
10^{11}\; \mbox{cm}^{-2}$ a resistivity of $\rho_c = 27\; k\Omega$ 
($= 1.05\; h/e^2$) is
observed, independent of the electric field.  By plotting $\rho_{xx}$ 
vs. $\delta_n/E^q$
for different values of the exponent and  $\delta_n =
|n-n_c|/n$, it is possible to find a $q$ for which all the
resistivity curves fall on one of two branches, one for the insulating
($\rho_{xx} > \rho_c$) and one for the metallic region ($\rho_{xx} < 
\rho_c$). 
An example of such a plot is shown in Fig. 2, for $q = 0.175$. In this 
figure the overlap has been optimized for the metallic 
phase.  Close to the critical point there is a symmetry between the 
two resistivity
branches, but this is lost at higher fields, where the insulator 
traces no longer
fall onto one scaling curve.

The transition can also be seen in the temperature dependence of 
$\rho_{xx}$. 
In Fig. 1c we show this for $T = 1.5\; K - 5\; K$.  The critical 
density, $n_c =
2.1\cdot 10^{11}\; \mbox{cm}^{-2}$, is very close to that found in 
the electric field
experiments.  For densities close to the critical point,
$d\rho_{xx}/dT$ changes sign as the temperature is lowered.  This 
makes the determination of
the critical density less accurate than in the E-field scaling.  For 
the temperature
dependence it is possible to obtain scaling, by plotting $\rho_{xx}$
vs.
$\delta_n/T^p$ where the exponent $p$ is chosen such that the 
measured traces fall onto
two separate scaling curves, with a reasonable symmetry between the 
two phases.

For a quantum phase transition the
resistivity is expected to be a function of $\delta_n/T^{1/z\nu}$ or
$\delta_n/E^{1/(z+1)\nu}$, depending on whether the correlation 
lengths are
controlled by the temperature or the electric field \cite{Sond}.
From the above stated $\rho _{xx}$ dependence on E and T we see that
the two exponents determined from E-field and temperature scaling,
depend on  the dynamical exponent $z$ and the correlation length 
exponent $\nu$: $q
= 1/(z+1)\nu$ and $p = 1/z\nu$. Assuming the same functional form of 
the resistivity for the E-field and temperature, the two types of 
scaling behaviour can be mapped onto each other via 
$T\rightarrow\kappa E^{z/(z+1)}$ with $\kappa$ being a system 
dependent constant.  For a strongly interacting 2D system $z = 1$ is 
expected \cite{Sond}. In order to test this we scale
$E^{1/2}$ and
$T$ on top of each other.  In Fig. 2 $\rho_{xx}$ is plotted against 
$\delta_n/T^{0.35}$ and
$\delta_n/(E^{1/2})^{0.35}$, with a reasonable scaling
behaviour for both curves.  In the inset $E^{1/2}$ has been 
multiplied by a
constant $\kappa = 0.4 K(m/V)^{1/2}$, which demonstrates the highly 
identical
resistivity dependence on $T$ and $E^{1/2}$ in the metallic regime, 
and strongly
supports $z = 1$.  This gives a value for
$\nu \approx 2.85$.  
\begin{figure} 
\begin{minipage}{0.48\linewidth}
\epsfig{file=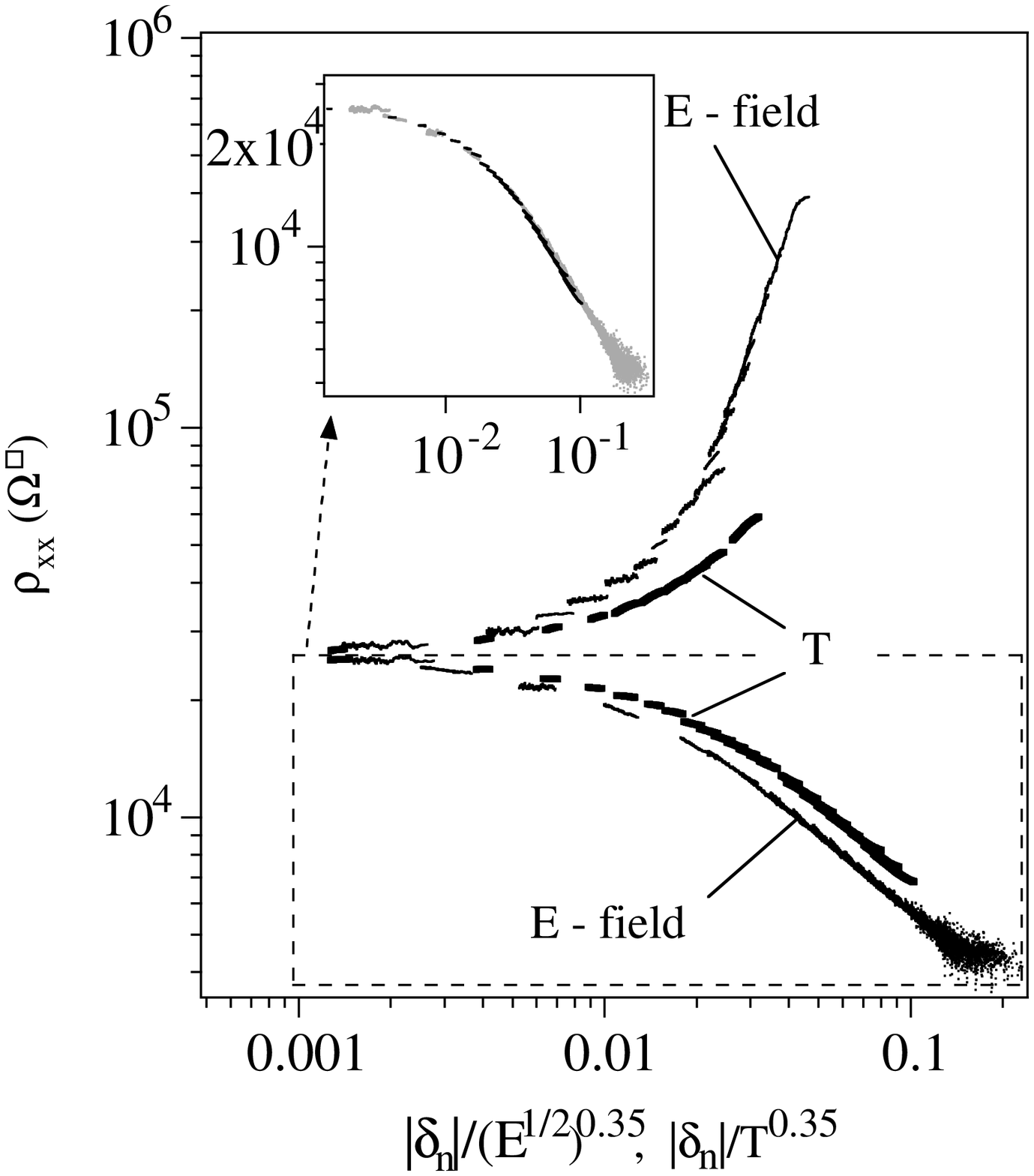,height=6.2cm}
\caption{Resistance vs. $|\delta_n|/T^{0.35}$ and $|\delta 
_n|/(E^{1/2})^{0.35}.$ Shown
in the inset is the metallic part of $\rho _{xx}\; vs.\;  |\delta 
_n|/T^{0.35}$ (black) and
$|\delta _n|/(\kappa E^{1/2})^{0.35}$ (grey), with $\kappa = 0.4$.}
\end{minipage}
\hfill
\begin{minipage}{0.48\linewidth}
\epsfig{file=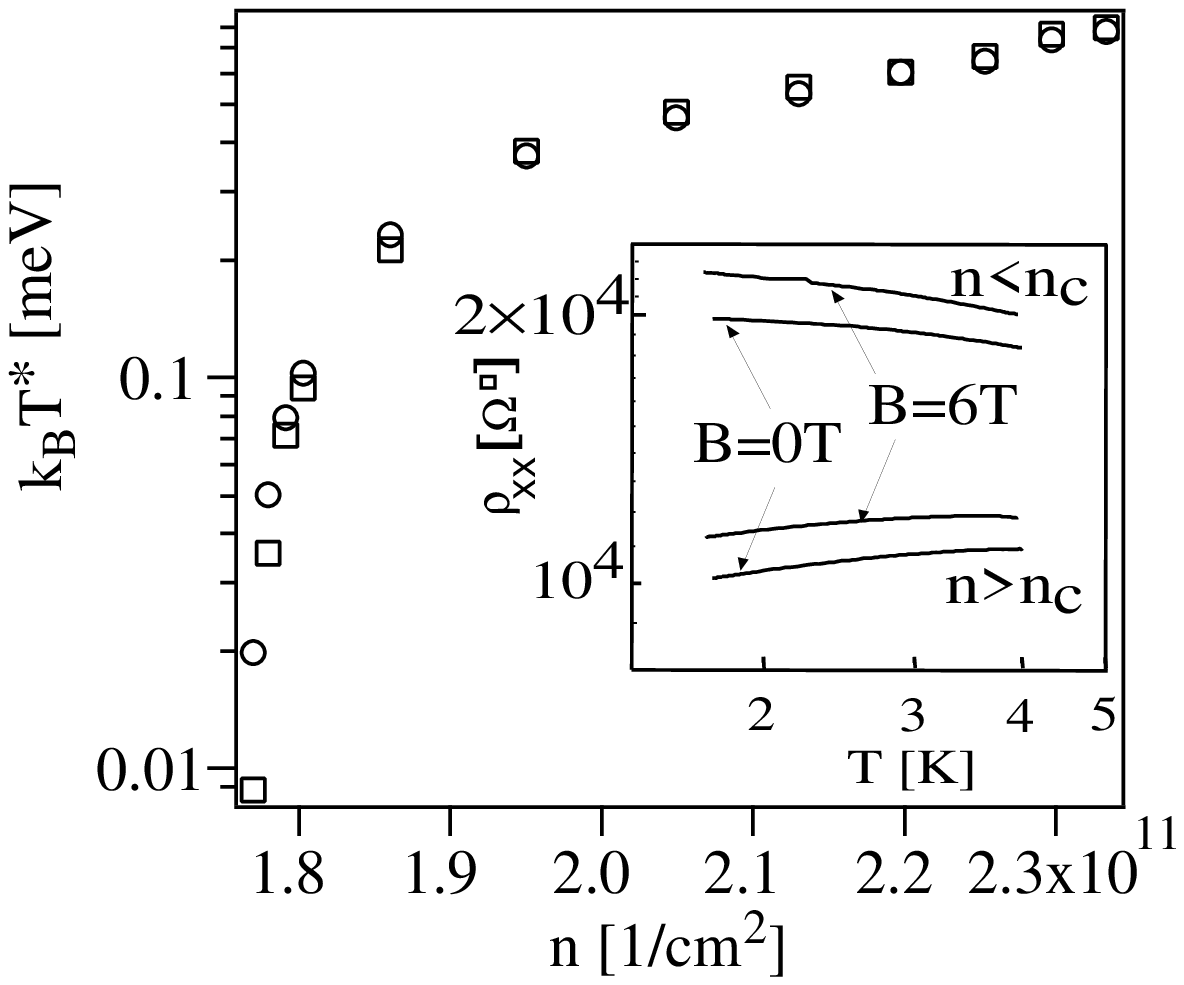,height=4.8cm}
\caption{The energy gap $k_BT^*$ as obtained from fits to the 
resistivity vs.
electric field curves (see text).  Squares: $B_{||} = 0T$, circles:  
$B_{||} = 4T$.   Inset:  Resistivity
vs. temperature for two different carrier densities, $n > n_c$ and $n 
< n_c$ for $B =
0T$ and $B_{||} = 6T$, respectively.}
\end{minipage}
\end{figure}
While $z$ is predicted to
be equal to 1 \cite{Sond}, we are not aware of any predictions for 
the value
of $\nu$ for the $B = 0$ 2D MIT. The $T\propto E^{1/2}$ dependence 
suggests that the elctric field can be interpreted in terms of 
quantum fluctuations \cite{Sond}, but we can not exclude the 
possibility of a heating mechansim due to the electric field which 
could lead to the same dependence.

In our system the transition is found for $k_Fl \approx 1$,
where $l$ is the mean free path as deduced from the mobility, and 
$k_F = 
\sqrt{2\pi n}$ is the Fermi wave vector.  This point also corresponds 
to a ratio
between the Coulomb interaction and the Fermi energy, $r_s = 4.9$.  
The value of
$r_s > 1$ in the metallic region has been considered as evidence for 
the
electron-electron interaction playing a major role \cite{Krav1}.

In contrast to the Si MOSFET material system \cite{Simo} in the 
present hole gases no resistivity increase with magnetic
field applied parallel to the 2DHG is observed, and the E-field 
behaviour remains the same.  This is quantified
by fitting the resistivity to $\rho_{xx} = \rho_0 + \rho_1 
exp{[-T^*/T]}$ \cite{Puda}, with $T = \kappa E^{1/2}$, where $\kappa 
= 0.4\; 
Km/V$ (see above), and  $\rho_0$ and $\rho_1$ as $T$--independent 
parameters.  $\rho_1$ and $T^*$ are used as fitting parameters, 
whereas
$\rho_0$ has been kept constant for all fits ($=  10 k\Omega$).  In 
Fig. 3, the
energy gap $k_BT^*$ vs. carrier density is shown for the case of zero 
field and a field $B_{||} = 4 T$ applied parallel to the SiGe quantum 
well.  There is no
indication of any magnetic quenching of the energy gap.  In the inset 
of Fig. 3,
$\rho_{xx}$ vs. $T$ is shown for $B = 0$ and for $B_{||} = 6 T$, for 
two different
carrier densities.  Except for an overall increase in the 
resistivity, the behaviour
remains essentially the same.  The difference between the magnetic 
quenching
of the transition in a Si MOSFET \cite{Simo} and the present 
observation can be understood
by considering the Zeeman effect in a SiGe 2 DHG.  The carriers 
populate a heavy
hole subband, with the characteristics $(J, m_J) = (3/2, \pm 3/2)$, 
where $J$ is
the total angular momentum.  Due to the symmetry of this state, $\bf 
J$ will be
aligned perpendicular to the quantum well, and this arrangement is 
reinforced
by the strain splitting, which further removes the lowest heavy hole 
state from the light
hole states.  Thus the total angular momentum feels an easy axis, and 
the
magnetic energy will be dominated by the magnetic field component
perpendicular to the quantum well (in order to break the LS coupling, 
a field of the order of $100 T$ would be needed).  The lack of any 
influence on the
metallic conduction by a parallel magnetic field, together with the 
breakdown of
this phase in systems without an easy axis, supports the hypothesis 
that a
spin related mechanism plays a key role for the MIT in 2 D.

In conclusion, $E$-field and temperature scaling behaviour have been
demonstrated for the MIT observed in a SiGe 2 DHG. 
Scaling plots lead to a dynamical exponent $z \approx 1$ and a 
correlation length
exponent $\nu \approx 2.85$.  A magnetic field up to $6 T$ applied 
parallel to the
interfaces does not quench the transition. 

{\small We wish to thank D. B\"{a}chle and T. Mezzacasa for assistance with 
the
processing, and P. Studerus for help with the electronics.  P. 
Coleridge is
gratefully acknowledged for helpful discussions.  This work was 
supported by the
Schweizerischer Nationalfonds and MINAST.\\}

\end{document}